\documentclass[draft]{agujournal2019}
\usepackage{float}
\usepackage{url}
\usepackage{soul}
\usepackage{amsmath}
\usepackage{amssymb}
\usepackage{ragged2e}
\usepackage{algorithm}
\usepackage{algpseudocode}
\usepackage{natbib}
\usepackage{booktabs}
\usepackage{graphicx}

\journalname{Geophysical Research Letters}

\makeatletter
\@fpsep=\textheight
\makeatother

\begin{document}
\justifying

\title{Breaking Waves Illuminate the Tidally Modulated Intertidal Zone through Fiber-Optic Seismology}

\authors{Haipeng Li\affil{1}, Hannah E. Glover\affil{2}, Meagan E. Wengrove\affil{2}, Fu Yin\affil{3}, Jingxiao Liu\affil{4}, Biondo L. Biondi\affil{1}}
\affiliation{1}{Department of Geophysics, Stanford University, Stanford, CA 94305, USA.}
\affiliation{2}{College of Engineering, School of Civil \& Construction Engineering, Oregon State University, Corvallis, OR 97331, USA.}
\affiliation{3}{Department of Earth, Environmental, and Planetary Sciences, Rice University, Houston, TX 77005, USA.}
\affiliation{4}{School of Civil and Environmental Engineering, Nanyang Technological University, Singapore.}

\correspondingauthor{Haipeng Li}{haipeng@sep.stanford.edu}

\begin{keypoints}
\item Breaking ocean waves provide a natural seismic source to image beach structure and monitor tidal responses with a buried fiber-optic array.
\item Imaging resolves softer foreshore and stiffer backshore sandy sediments to 6 m depth, reflecting pore-water and grain-contact controls.
\item Tidal flooding and drainage modulate attenuation due to partial saturation and alter fluid--solid boundaries for elastic wave propagation.
\end{keypoints}

\begin{abstract}
Tidal stage and water content alter the geotechnical properties of beach sediments, yet these effects remain difficult to observe beneath energetic surf. Here, we use breaking waves as a natural seismic source to image beach structure and sediment properties. Seismic interferometry on a cross-shore Distributed Acoustic Sensing array at Rockaway Beach, Oregon, reveals foreshore softening and backshore stiffening, driven by pore-water control of effective stress and capillary forces, with additional packing and compaction effects. Intertidal effective attenuation increases at low tide and exhibits flood--ebb hysteresis due to partial saturation during drainage. Inundation also converts Rayleigh waves on exposed ground into Scholte waves that shoal along the submerged interface toward the shoreline. Phase velocity decreases with increasing water depth, consistent with acoustic--elastic wave modeling. Our study advances imaging and monitoring of mechanical and hydrological states in the beach subsurface, an underexplored frontier between ocean dynamics and the solid Earth.
\end{abstract}

\section*{Plain Language Summary}
Tides repeatedly flood and drain beaches, changing subsurface water content, but these processes are difficult to observe beneath breaking surf. We use a buried fiber-optic cable at Rockaway Beach, Oregon, to record ground vibrations generated by breaking waves. The recordings image beach sand and reveal softer sand toward the ocean and stiffer sand landward. We attribute this contrast mainly to pore water changing the forces between sand grains, with additional effects from packing and compaction. Seismic energy is more strongly attenuated at low tide, with distinct attenuation paths during rising and falling tides at the same water levels. This hysteresis indicates that drainage and partial saturation continue to influence the sand as tides change. Flooding also converts waves traveling along exposed ground into Scholte waves along the submerged seabed. These waves shoal toward the shoreline and slow as water depth increases, consistent with wave simulations. Breaking waves and fiber-optic sensing can therefore image beach structure and track groundwater-related changes across the boundary between ocean and land.

\newpage

\section{Introduction}
\begingroup
Sandy beaches constitute a large fraction of the world's coastlines \citep{luijendijk2018state} and form a physical boundary between ocean forcing and landward environments, governing sediment transport \citep{wright1984morphodynamic}, aquifer interactions \citep{robinson2007tide, xin2010effects, heiss2014saltwater}, and shoreline evolution \citep{barnard2017extreme, vousdoukas2020sandy}. In intertidal environments, tides and breaking waves drive infiltration and drainage \citep{nielsen1990tidal}, continually altering pore-water distribution and effective stress within the sand. Shoreline migration produces spatially variable responses to flooding and drainage across beach faces \citep{raubenheimer1999tidal, heiss2015coupled}. These responses connect coastal groundwater exchange with sediment mechanics \citep{turner1998swash}, yet remain difficult to observe beneath energetic surf. Consequently, coastal geological structure and its interactions with hydrological processes remain poorly constrained across the land--sea transition \citep{weymer2022coastal}.

Resolving these dynamic processes requires measurements connecting cross-shore sediment structure with tidal infiltration and drainage \citep{heiss2015coupled}, which has been difficult with existing observational techniques. Remote sensing and lidar can characterize beach morphology and shoreline dynamics \citep{pardo2012automatic, vos2019coastsat}. In-situ sensors record local water-level and pore-pressure fluctuations \citep{raubenheimer1999tidal, turner1997rapid}, while geoelectrical methods map pore-fluid salinity and saturation \citep{binley2015emergence, costall2018electrical, mcdonnell2023four}. However, integrating these datasets to examine dynamic pore-water variability is challenging. Seismic methods link subsurface structure to hydrological processes through shear-wave velocity ($V_s$), which increases with sediment grain-frame stiffness and also depends on bulk density. In unconsolidated sand, grain-frame stiffness is governed by forces at grain contacts and therefore responds to pore-water conditions through effective stress and capillary forces \citep{hardin1972shear, zimmer2007seismic, lu2010closed}. Active surveys map $V_s$ beneath exposed and submerged ground \citep{long2020multichannel}, while passive monitoring tracks groundwater-related velocity changes in terrestrial aquifers \citep{clements2018tracking, mao2022space, li2025near, li2025fiber}. Recent DAS observations also resolve terrestrial soil responses to wetting and drying through capillary effects \citep{shi2026agroseismology}. However, active surf and a tidally migrating shoreline make continuous seismic monitoring across beaches challenging and this environment remains understudied.

Ocean waves transfer mechanical energy into the solid Earth and contribute to Earth's ambient seismic wavefield. At low frequencies, this energy produces microseisms \citep{longuet1950theory, Webb1998, ardhuin2015ocean}, which have been used to image crustal structure through ambient-noise tomography \citep{shapiro2005high}. In the surf zone, breaking waves dissipate energy across the beach face, transferring part into high-frequency seismic waves \citep{young2013ground, poppeliers2015high, thompson2022seismic}. These signals provide a local source for probing beach sediments through successive tidal cycles. Utilizing them requires dense measurements under energetic surf. Distributed Acoustic Sensing (DAS) transforms fiber-optic cables into dense arrays of strain-rate (or strain) sensors, enabling continuous measurements in harsh environments \citep{zhan2020distributed, lindsey2021fiber, liu2025urban}. Seafloor fibers record microseisms \citep{lindsey2019illuminating, sladen2019distributed, williams2019distributed} and oceanographic signals from internal waves and tides \citep{williams2023fiber}, while nearshore deployments track waves, tides, and currents \citep{hu2024submarine, song2024near, glover2024measuring, glover2026comparisons}. Beach arrays resolve tidal modulation of wave-to-seismic energy conversion \citep{ko2026tide}, and seafloor DAS interferometry recovers Scholte-wave dispersion for imaging shallow sediments \citep{cheng2021utilizing}. Despite these advances, fiber-optic seismic imaging of beach structure and monitoring of subsurface responses to tidal flooding and drainage remain largely unexplored.

Here, we use a cross-shore DAS array buried at Rockaway Beach, Oregon, to investigate beach structure and its tidal response using breaker-generated seismic waves. Throughout this study, waves denote seismic waves, whereas breaking waves denote surface-water motion. Seismic interferometry and dispersion inversion characterize beach structure by resolving a $V_s$ section across the beach face, revealing cross-shore stiffness variations within the sandy coastal aquifer that are controlled by hydrological conditions. We combine time-lapse attenuation analysis with acoustic--elastic modeling to distinguish hydrological changes within the sediments from water-column effects on seismic-wave propagation. Using breaking waves as a natural seismic source, our study reveals how coastal hydrodynamics interact with sediment processes beneath the beach, an underexplored frontier at the land--sea transition.

\par\endgroup
\section{Data and Methods}

\subsection{Field Experiment and Data}

\begin{figure}[tbp]
\centering
\includegraphics[width=1.0\textwidth]{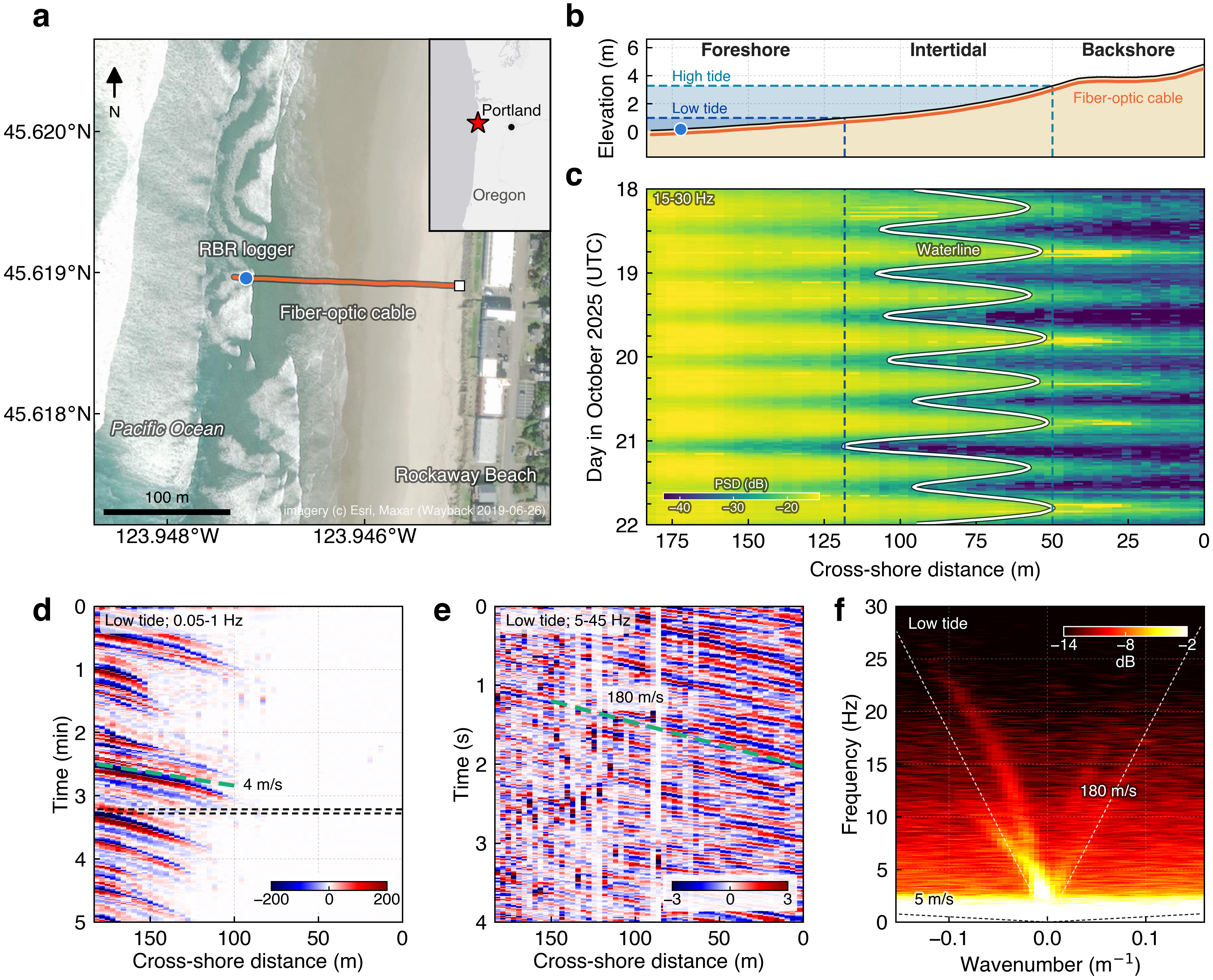}
\caption{Cross-shore DAS deployment and tidal modulation of the recorded wavefield at Rockaway Beach, Oregon. (a) Buried fiber-optic cable (orange) and RBR pressure sensor (blue circle) overlaid on aerial imagery; the inset shows the study site. (b) Surveyed beach profile with the cable buried approximately $0.30~\text{m}$ below the surface. Dashed lines mark the minimum and maximum still-water levels ($\eta = 1.0$ and $3.3~\text{m}$), defining the submerged foreshore, intertidal zone, and backshore. (c) Spatiotemporal variations in DAS wavefield power within the $15\text{--}30~\text{Hz}$ band and the moving waterline (white). (d) Five-minute hydrodynamic-band record during low tide. The green dashed line indicates $4~\text{m/s}$, and black dashed lines bracket the interval in (e). (e) Four-second, trace-normalized seismic-band wavefield. (f) Frequency--wavenumber spectrum of the corresponding low-tide record, with reference velocities of $180~\text{m/s}$ (white) and $5~\text{m/s}$ (black).}
\label{fig:1}
\end{figure}

We collected DAS data at Rockaway Beach, Oregon, an open sandy coast exposed to the North Pacific and semidiurnal tides (Figure~\ref{fig:1}a). A cross-shore cable was trenched approximately $0.30~\text{m}$ into the sand; DAS data were recorded with a Sintela Onyx 1.0 interrogator at $1,000~\text{Hz}$, with a gauge length and channel spacing of $3.19~\text{m}$. Surveys and tap tests geolocated 58 analyzed channels spanning approximately $180~\text{m}$ from the backshore to the low-tide terrace, with cross-shore distance $x$ increasing seaward. Surveyed elevations descend from $4.80$ to $0.09~\text{m}$ (NAVD88; Figure~\ref{fig:1}b). An RBR Duet pressure and temperature sensor sampling at $4~\text{Hz}$ measured water depth, which was converted to still-water elevation $\eta$ on the same vertical datum (Text S1).

The analysis covers 18--21 October 2025 (UTC), when $\eta$ ranged from $0.98$ to $3.29~\text{m}$. These limits define a continuously submerged foreshore (20 channels, $x>118~\text{m}$), an intermittently inundated intertidal zone (22 channels, $50\le x\le118~\text{m}$), and a backshore above the maximum still-water level (16 channels, $x<50~\text{m}$; Figure~\ref{fig:1}b). These zones describe inundation relative to still-water level rather than individual swash events.

\subsection{Breaking-Wave Seismology}

\begingroup
Breaking waves provide a recurring seismic source for imaging beach structure and monitoring its tidal response. To assess these signals, we distinguish direct hydrodynamic loading from propagating seismic waves. Power spectral densities (PSDs) characterize frequency content across the beach and tidal cycle. We use the mean PSD within $15\text{--}30~\text{Hz}$, averaged in decibels and here termed band power, to track spatial and tidal variations. Hydrodynamic and seismic bands are analyzed in the frequency--wavenumber ($f\text{--}k$) domain (Text S2).

To image beach subsurface structure, we retrieve virtual-source gathers (VSGs) by cross-correlating DAS recordings and stacking over four days \citep{bensen2007processing}. The stronger landward-propagating branch provides dispersion measurements over apertures up to $90~\text{m}$. Fundamental-mode ($5\text{--}12~\text{Hz}$) and second-higher-mode ($12\text{--}28~\text{Hz}$) curves are inverted for local $V_s$ profiles using layered elastic models with an overlying water layer where submerged. With water-layer thickness fixed at the analysis-period mean inundation depth, the profiles provide a tidally averaged structural baseline (Text S2).

Beyond characterizing beach subsurface structure, we track the dynamic response of sandy sediments to tidal flooding and drainage using time-lapse seismic attenuation, quantified by $1/Q$, the inverse of the seismic quality factor. Pore-water redistribution can alter wave-induced fluid-flow dissipation, making attenuation sensitive to hydrological conditions \citep{pride2004seismic, mavko2020rock}. We estimate $1/Q$ from the frequency-dependent amplitude decay of landward-propagating wavefields using the spectral-ratio method \citep{tonn1991determination, yuan2024using}. Because the measured amplitude decay reflects source strength, cable coupling, scattering, and inundation-related propagation in addition to energy dissipation within the sediment, we interpret $1/Q$ as effective attenuation. Effective attenuation represents the combined amplitude decay observed in the DAS correlations, whereas intrinsic attenuation refers only to energy dissipated within the sediment \citep{lin2011reliability}.

Tidal flooding changes the boundary conditions for seismic-wave propagation. Exposed sediment is bounded by air and behaves as a traction-free surface that supports Rayleigh waves, whereas submerged sediment is coupled to the water through a fluid--solid interface that supports Scholte waves \citep{carcione2018waves}. As the shoreline moves, Scholte waves shoal along this dynamic interface, and their phase velocity varies with water depth. To distinguish water-depth effects from sediment changes, we track foreshore phase velocities and use coupled acoustic--elastic wave modeling to simulate Scholte-wave dispersion and shoaling while varying only water depth. Spectral-element simulations compare low- and high-tide wavefields \citep{komatitsch1999introduction, tromp2008spectral}. This approach separates water-layer effects from sediment changes when interpreting tidal variations in seismic phase velocity (Text S3).
\par\endgroup

\begin{figure}[tbp]
\centering
\includegraphics[width=1.0\textwidth]{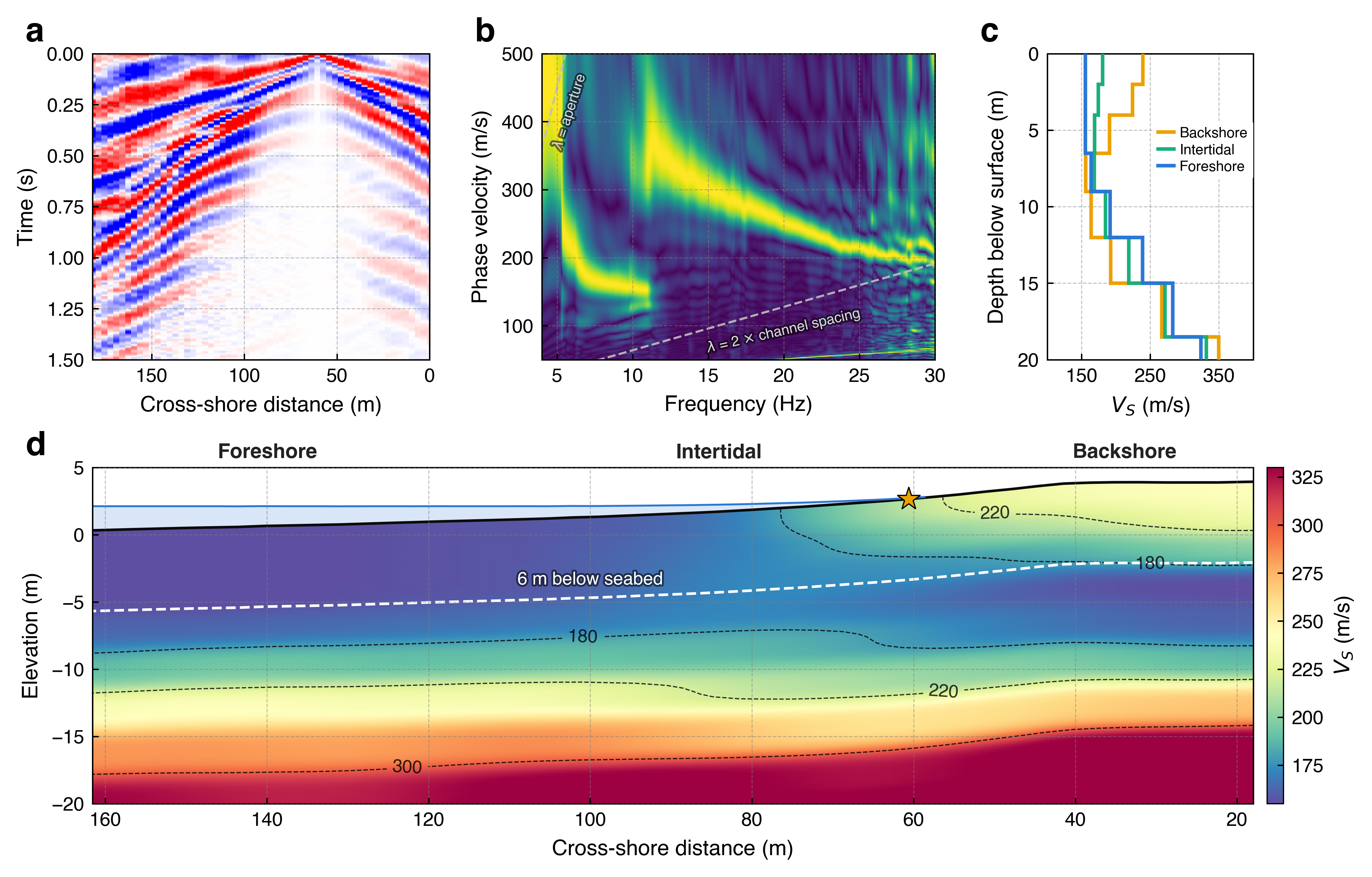}
\caption{Beach subsurface structure imaged using breaking-wave seismic energy. (a) Virtual-source gather (VSG) for a source in the intertidal zone at $x = 61~\text{m}$ (star in d). (b) Corresponding dispersion spectrum showing the fundamental and second higher modes; dashed lines mark the aperture and spatial-sampling constraints on the usable frequency range. (c) Zone-averaged $V_s$ profiles beneath the backshore (yellow), intertidal zone (green), and submerged foreshore (blue), with depth measured below the local sediment surface. (d) Inverted cross-shore $V_s$ section beneath the surveyed topography using tidal-average water depth. The white dashed line marks the upper $6~\text{m}$ examined in Figure~\ref{fig:3}b.}
\label{fig:2}
\end{figure}

\section{Results}
\subsection{Breaking Waves as Seismic Sources}

\begingroup
The recorded DAS wavefield changes across the beach as tidal flooding and drainage shift the waterline. The landward edge of the high-power region tracks this movement through successive tidal cycles (Fig.~\ref{fig:1}c ), spanning approximately $68~\text{m}$ across the intertidal zone over a $2.3~\text{m}$ tidal range. At channels that alternate between inundation and exposure, $15\text{--}30~\text{Hz}$ band power is approximately $9~\text{dB}$ higher during inundation. On exposed channels, power in this band decreases landward by approximately $1.0~\text{dB}$ per $10~\text{m}$, and the spectra show a peak near $3~\text{Hz}$ and reduced power above $10~\text{Hz}$, whereas inundated channels show broadband power extending to $100~\text{Hz}$ (Figure~S2). 

These spectral contrasts are accompanied by two distinct propagation regimes. Low-frequency signals below $1~\text{Hz}$ cross inundated channels at approximately $4\text{--}5~\text{m/s}$, consistent with hydrodynamic loading by shallow-water surface gravity waves (Figure~\ref{fig:1}d). Higher-frequency seismic waves cross both submerged and exposed sand at approximately $180~\text{m/s}$ (Figure~\ref{fig:1}e). In the $f$--$k$ domain, these arrivals form a fundamental mode at $5\text{--}10~\text{Hz}$ and a faster higher mode extending to approximately $30~\text{Hz}$, with high-frequency sensitivity limited by spatial sampling and gauge averaging (Figure~\ref{fig:1}f). These seismic waves propagate in both directions, with a stronger landward branch at low tide and a seaward branch of comparable strength at high tide (Figure~S1). This seaward propagation could reflect waves returning from shore or local generation by breakers over the array, although propagation direction alone cannot distinguish these origins. The coherent landward-propagating waves provide the signals used for interferometric imaging.
\par\endgroup

\begingroup
\subsection{Beach Structure and Spatial Variations in Sediment Stiffness}
\par\endgroup

\begin{figure}[tbp]
\centering
\includegraphics[width=1.0\textwidth]{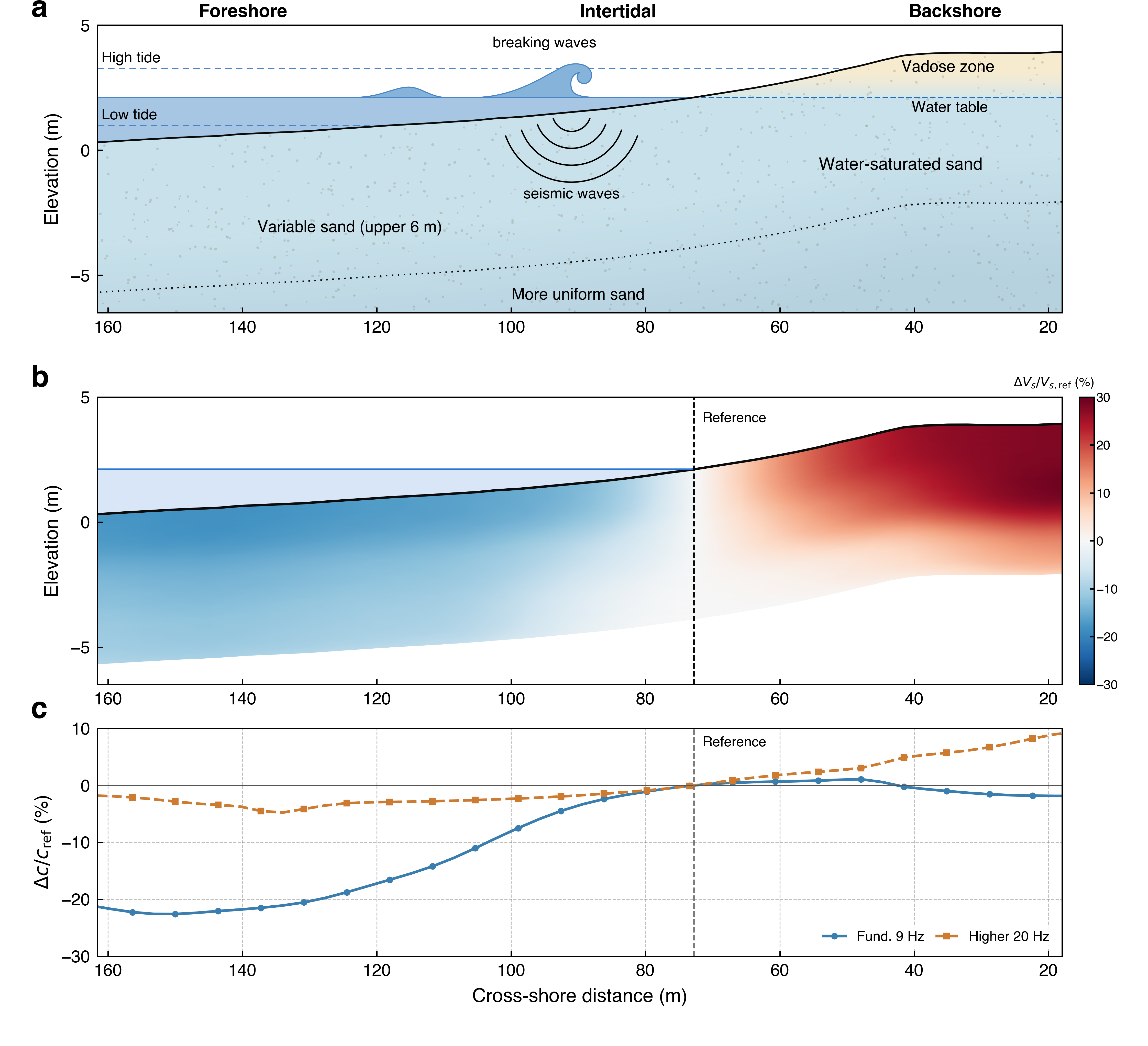}
\caption{Cross-shore sediment stiffness and dispersion sensitivity. (a) Conceptual beach section showing the vadose zone, water table at mean still-water elevation, and water-saturated sand; the dotted line marks $6~\text{m}$ depth. (b) Relative $V_s$ anomalies referenced to mean profiles within $5~\text{m}$ of the mean-water shoreline (vertical dashed line) at equal depth below the local sediment surface. Blue and red indicate lower and higher velocities. (c) Phase-velocity anomalies for the $9~\text{Hz}$ fundamental mode (blue circles) and $20~\text{Hz}$ second higher mode (orange squares). Here, $\Delta c/c_{\mathrm{ref}}=(c-c_{\mathrm{ref}})/c_{\mathrm{ref}}$, where $c_{\mathrm{ref}}$ is the mode-specific phase velocity at the dashed shoreline reference.}
\label{fig:3}
\end{figure}

\begingroup
Using breaking waves as an energetic seismic source, we retrieve VSGs with coherent surface-wave arrivals (Figure~\ref{fig:2}a). Their dispersion spectra resolve the fundamental and second higher modes (Figure~\ref{fig:2}b), and inverting the corresponding dispersion curves yields a cross-shore $V_s$ section. Representative fits to the dispersion inputs are shown in Figure~S4b--d. This hydrogeological section shows higher velocities in the shallow subsurface beneath the backshore than beneath the submerged foreshore, with the strongest lateral contrasts near the sediment surface (Figure~\ref{fig:2}c, d). In the upper approximately $6~\text{m}$, foreshore velocities are approximately $19\%$ to $35\%$ below backshore values. Lateral contrasts are markedly weaker at $6.5\text{--}9~\text{m}$ depth. Velocities increase toward a higher-velocity unit below around $18~\text{m}$ and exceed $300~\text{m/s}$ at $20~\text{m}$. Within this shallow zone, $V_s$ primarily reflects grain-frame stiffness, and we interpret the lower foreshore values as relative softening and the higher backshore values as relative stiffening (Figure~\ref{fig:3}a) \citep{mindlin1949compliance, dvorkin1996elasticity, zimmer2007seismic, mavko2020rock}.
\par\endgroup

\begingroup
Figure~\ref{fig:3}c compares the well-resolved $9~\text{Hz}$ fundamental and $20~\text{Hz}$ second-higher modes. Each mode is normalized to its own phase velocity at the same shoreline position, so both anomalies are zero there. Relative to the shoreline reference, the $20~\text{Hz}$ higher-mode phase velocity increases landward by up to approximately $9\%$ across the backshore, whereas the $9~\text{Hz}$ fundamental-mode phase velocity remains approximately constant. Because the higher mode is more sensitive to shallow sediments while the fundamental mode integrates over a broader and deeper interval (Figure~S6), this mode-dependent response indicates that backshore stiffening is concentrated near the surface and weakens with depth. Across the foreshore, both modes fall below the shoreline reference, with reductions of approximately $20\%$ for the fundamental mode and a few percent for the higher mode. These phase-velocity changes also include the influence of increasing water depth. After accounting for the water layer in the inversion, Figure~\ref{fig:3}b shows that lower $V_s$ persists through the upper approximately $6~\text{m}$ and is strongest near the seabed, indicating that foreshore softening is concentrated in shallow sediments but extends several meters downward.
\par\endgroup

\begingroup
We interpret this cross-shore contrast in terms of effective stress and grain-contact conditions shaped by pore-water distribution and sediment packing. Higher $V_s$ in the shallow backshore indicates increased grain-frame stiffness above the water table. Capillary suction can strengthen grain contacts in partially saturated sand, whereas buoyancy reduces effective overburden beneath the submerged foreshore. In the submerged foreshore, hydrostatic loading alone cannot explain the observed velocity reduction because it increases total stress and pore pressure equally \citep{terzaghi1996soil, zimmer2007seismic, lu2010closed, nielsen1990tidal, raubenheimer1999tidal}. Variations in packing, porosity, and compaction history can further modify sediment stiffness \citep{zimmer2007sorting, mavko2020rock}, while pore-pressure fluctuations induced by ocean surface gravity waves alter the effective stress carried by grain contacts \citep{madsen1978wave, yamamoto1978response}. Wave-driven sediment reworking can also modify grain packing and contact geometry in shallow beach sediments \citep{ciavola1997field}. These processes explain the observed mechanical contrast between the stiffer backshore and softer foreshore.
\par\endgroup

\subsection{Tidal Modulation of Seismic Attenuation and Wave Propagation}

\begin{figure}[tbp]
\centering
\includegraphics[width=1.0\textwidth]{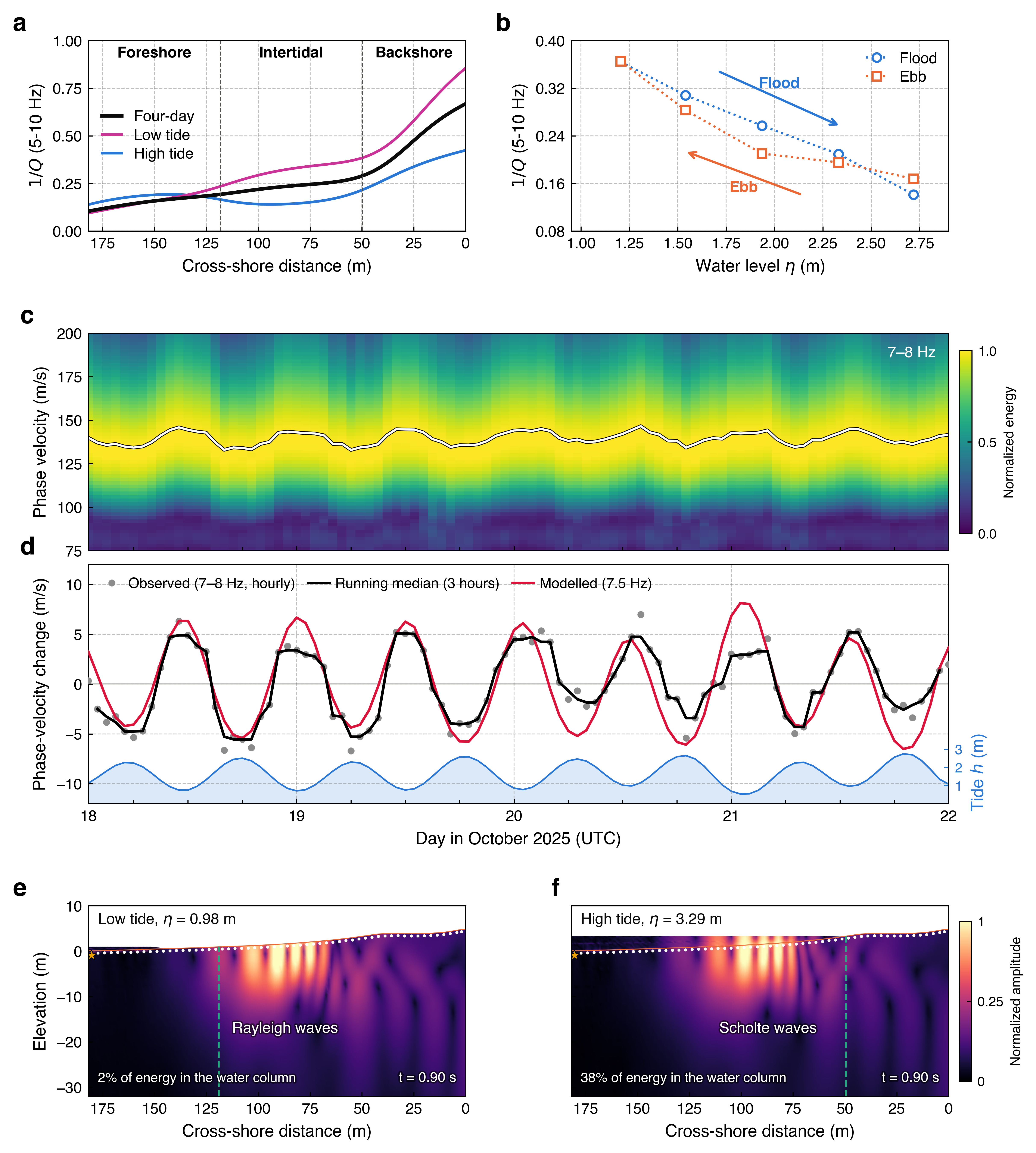}
\caption{Tidal variations in effective attenuation and acoustic--elastic wave propagation. (a) Effective attenuation ($1/Q$, $5\text{--}10~\text{Hz}$) along the array at high tide (blue), low tide (magenta), and averaged over time (black). (b) Median intertidal $1/Q$ during flood (blue circles) and ebb (orange squares) across five water-level intervals; arrows indicate tidal direction. (c) $7\text{--}8~\text{Hz}$ dispersion spectra with tracked phase velocities (white). (d) Observed (gray), smoothed (black), and modeled (red) phase-velocity changes, with water depth in blue. (e, f) Simulated low- and high-tide wavefields through the $V_s$ section. Stars mark the source, dashed lines the waterlines, and dots the DAS channels; percentages indicate the modeled water-column energy fraction.}
\label{fig:4}
\end{figure}

\begingroup
Tidal stage modulates seismic attenuation across the intertidal zone and adjacent backshore by altering water depth and saturation. Median $1/Q$ increases from $0.15$ at high tide to $0.33$ at low tide, whereas attenuation across the continuously submerged foreshore remains nearly constant (Figure~\ref{fig:4}a). Across five water-level intervals, the intertidal median $1/Q$ decreases with rising water level (Figure~\ref{fig:4}b). Flood and ebb tides trace distinct attenuation paths at equivalent water levels, forming a hysteresis loop. During flood tide, shoreline inundation advances faster than the groundwater table rises, whereas during ebb tide, groundwater drainage lags shoreline retreat \citep{nielsen1990tidal, raubenheimer1999tidal}. This asymmetric hydrological response leaves sediments at the same water level in different saturation states, producing the observed attenuation hysteresis. Continued drainage toward low tide allows air to enter the pore space and partially displace pore water, producing patchy saturation. This partially saturated state attenuates seismic waves more strongly than the more fully saturated state at high tide. Under patchy saturation, seismic loading generates local pore-pressure gradients between regions with contrasting fluid content, driving wave-induced fluid flow that dissipates seismic energy and increases attenuation \citep{white1975computed, pride2004seismic, mavko2020rock}. This hysteresis likely also influences the mobility of the sediment, suggesting that models of beach sediment transport should account for tidal lag in addition to current tidal stage.
\par\endgroup

\begingroup
Tides produce a distinct response on the submerged foreshore by changing the overlying water layer. Tracked Scholte-wave phase velocities over $7\text{--}8~\text{Hz}$ vary by about $10~\text{m/s}$ in antiphase with water level, with a correlation of $-0.85$ (Figure~\ref{fig:4}c, d; Figure S8). The measured water-depth sensitivities are approximately $-10~\text{m/s}$ per meter over $6\text{--}7~\text{Hz}$ and $-4.8~\text{m/s}$ per meter over $7\text{--}8~\text{Hz}$, compared with modeled values of $-9$ and $-6.7~\text{m/s}$ per meter at the respective band centers. Acoustic--elastic simulations capture the observed decrease in phase velocity by varying only water depth. They also show a broader transition from Rayleigh waves along the exposed beach to Scholte waves along the submerged sediment--water interface (Figure~\ref{fig:4}e, f). From low tide ($\eta = 0.98~\text{m}$) to high tide ($\eta = 3.29~\text{m}$), the fraction of modeled seismic-wave energy concentrated in the water column increases from approximately $2\%$ to $38\%$ (see also Figure~S7 and Movie S1; Text S3). This energy redistribution and Scholte-wave shoaling explain the delayed high-tide arrivals, showing that inundation alters the foreshore wavefield even at fixed sediment stiffness. These results identify changing inundation as a control on seismic propagation that must be distinguished from changes in sediment properties.
\par\endgroup

\begingroup
These time-lapse observations reveal coupled hydrological and elastodynamic responses at the dynamic boundary between the ocean and solid Earth. The hydrological response reflects changing groundwater saturation, which modulates the effective attenuation of seismic waves across the intertidal zone and adjacent backshore. The elastodynamic response reflects changing water depth, which controls the phase velocity of seismic waves and redistributes wave motion across the submerged foreshore. Separating subsurface hydrological changes from water-column propagation effects enables coastal seismic observations to track tidal flooding, drainage, and groundwater dynamics. 
\par\endgroup

\section{Discussion and Conclusions}

\begingroup
Here we demonstrate that a cross-shore DAS array provides repeated seismic illumination across the moving shoreline, resolving beach structure and tidal variability from the backshore to the submerged foreshore. Combining dispersion inversion, attenuation analysis, and acoustic--elastic modeling separates sediment structure and hydrological change from water-column propagation effects. These results show that cost-effective DAS monitoring links the mechanical, hydrological, and wave-propagation responses of sandy beaches.
\par\endgroup

\begingroup
The $V_s$-derived hydrogeological section and time-lapse attenuation provide complementary constraints on the structure and hydrological dynamics of the sandy coastal aquifer. Cross-shore variations in $V_s$ reflect the aquifer's hydrological state and sediment packing shaped by compaction history and wave-driven reworking. Attenuation is sensitive to pore-water redistribution, revealing distinct drainage and infiltration states at the same sea level. The resulting flood--ebb attenuation hysteresis offers a seismic constraint on the lag between ocean forcing and the beach water table and, when integrated with groundwater observations and flow modeling, informs aquifer hydraulic properties \citep{nielsen1990tidal, raubenheimer1999tidal, abarca2013transient, heiss2015coupled}. Rayleigh--Scholte conversion provides a separate measure of inundation and its effect on wave propagation. Accounting for this effect prevents velocity changes caused by the overlying water column from being misinterpreted as changes in sediment properties. These observations link groundwater dynamics with sediment mechanics, a connection relevant to coastal morphodynamics because infiltration and exfiltration alter sediment mobility through pore-pressure gradients and near-bed flow \citep{turner1997rapid, turner1998swash}.
\par\endgroup

\begingroup
Our case study shows the potential of DAS monitoring to track beach evolution beyond individual tidal cycles, including storm-driven erosion, recovery, and seasonal hydrological change. Quantitative interpretation still requires distinguishing sediment dissipation from source, coupling, scattering, and propagation effects within the effective attenuation measurement. Co-located pore-pressure and moisture sensors, together with wave-propagation modeling, provide independent constraints on the hydrological sensitivity of effective attenuation. An additional advantage of DAS is its complementarity with electromagnetic methods, because seismic observations primarily constrain saturation-related sediment mechanics, whereas electrical conductivity responds strongly to both saturation and pore-water salinity \citep{binley2015emergence, grana2026rock}. The hydrogeological section is a tidally averaged $V_s$ model derived over apertures up to $90~\text{m}$. Shorter apertures and full-waveform inversion improve spatial localization \citep{li2025fiber, eppinger2024near}, while coupling groundwater-flow and sediment-mechanics models with rock physics links seismic changes to water storage and drainage timescales \citep{abarca2013transient, heiss2015coupled, grana2026rock}.
\par\endgroup

\begingroup
We show that breaking-wave seismology with passive DAS images a sandy beach and monitors its tidal response across the land--sea transition. The measurements resolve cross-shore variations in sediment stiffness, flood--ebb hysteresis in intertidal effective attenuation, and Scholte-wave slowing with increasing water depth, consistent with acoustic--elastic modeling. By combining fiber-optic sensing with breaking waves as persistent natural seismic sources, this study establishes a scalable framework for investigating underexplored coastal environments where ocean dynamics and solid-Earth processes interact. Because sandy beaches are widespread globally, this approach supports assessments of coastal vulnerability and resilience to El Ni\~no-driven wave extremes and other coastal storms.
\par\endgroup

\section*{Open Research Section}
The ambient-noise cross-correlation code used for seismic interferometry is publicly available at \texttt{https://github.com/ohmyshake/noisecc}. The supporting data and code are available on Zenodo under a CC BY 4.0 license at \url{https://doi.org/10.5281/zenodo.24120513} \citep{li2026intertidal}.

\section*{Author Contributions}
H.L., H.E.G., M.E.W., and B.L.B. conceived the study. H.E.G. and M.E.W. designed and led the fieldwork and provided coastal-process interpretation. H.L., F.Y., and J.L. developed the seismic analysis code and workflows. H.L. processed and analyzed the DAS data, prepared the figures, and drafted the manuscript. B.L.B. supervised the project. All authors discussed the results and revised the manuscript. The authors declare no competing interests.

\acknowledgments
H.L. and B.L.B. were supported by the Stanford Earth Imaging Project (SEP) affiliate companies. H.L. was supported by the Jon Claerbout Fellowship. H.E.G. was supported by the National Science Foundation under Award No. 2407466 (OCE-PRF), which also funded the data collection.

\nocite{castagna1985rock,gardner1974formation,herrmann2013computer,luu2021disba,park1998imaging}
\bibliography{ref}

\end{document}